\documentstyle[12pt,epsf]{article}  

\advance\hoffset by -7mm

\makeatletter
\@addtoreset{equation}{section}
\makeatother
\renewcommand{\title}[1]{\null 

\noindent{\Large{\bf #1}}\vspace{10mm}

\noindent {\large By }}
\newcommand{\authors}[1]{\noindent{\large #1}\vspace{3mm}

}
\newcommand{\address}[1]{\noindent #1\vspace{5mm}

}
\renewcommand{\abstract}[1]{\vspace{9mm}

\noindent{\small{\em Abstract.} #1}\vspace{2mm}

}     

\def\be{\begin{equation}}
\def\ee{\end{equation}} \def\eel#1 {\label{#1}\end{equation}}
\def\rz#1 {(\ref{#1}) } \def\ry#1 {(\ref{#1})}

   \let\m=\mu
\let\n=\nu    \let\s=\sigma

\let\O=\Omega \let\S=\Sigma  
  
\def\CP{{\cal P}} \def\CM{{\cal M}} \def\CN{{\cal N}} 
\def\0{\over } \def\1{\vec } \def\2{{1\over 2}} \def\4{{1\over 4}}
\def\5{\bar } \def\6{\partial }
\def\({\left(} \def\){\right)} \def\<{\langle } \def\>{\rangle }
\def\[{\left[} \def\]{\right]}

\newcommand{\dR}{\mbox{{\rm I \hspace{-0.86em} R}}}

\begin{document}

\hspace*{\fill} \vspace*{2mm}
                 TUW - 96 - 14\\
\hspace*{\fill}  PITHA - 96/23\\

\title{Classical and Quantum Aspects of 1+1 Gravity}
\authors{Thomas Kl\"osch%
\footnote{supported by Fonds zur F\"orderung der wissenschaftlichen Forschung
   (FWF), project P10221-PHY.},
         Peter Schaller%
\footnote{talk delivered at Journees Relativistes `96 (Ascona, Switzerland).},}
\address{Institut f\"ur Theoretische Physik, Technische Universit\"at Wien,\\
A-1040 Vienna, Austria}
\authors{and Thomas Strobl}
\address{Institut f\"ur Theoretische Physik, RWTH-Aachen,\\
D52056 Aachen, Germany}
\abstract{We present a classification of all global solutions (with Lorentzian
 signature) for any general 2D dilaton gravity model. For generic choices of
 potential-like terms
 in the Lagrangian one obtains maximally extended solutions on
 arbitrary non-compact two-manifolds, including various black-hole
 and kink configurations.  We determine all physical quantum
 states in a Dirac approach. In some cases the spectrum of the (black-hole)
 mass operator is found to be sensitive to the signature of the theory,
 which may be relevant in view of current attempts to implement a
 generalized Wick-rotation in 4D quantum gravity.}

There is some good news and some bad news about the subject of this talk.
To start with the bad news:
I am going to speak about {\it pure} gravity-Yang-Mills systems. 
This means that matter fields are not included (with the exception of dilaton
fields, which are not regarded as matter fields in our context). Furthermore,
I am going to speak about $(1+1)$ dimensional theories. As we seemingly live
in a $(3+1)$ dimensional universe, this might not be precisely what one
eventually is interested in. 

And here is the good news: We can give a comprehensive treatment of
the subject, encompassing a large variety of different gravity
models. We deduce a method to calculate all classical solutions of all
of these models on the local as well as on the global level. And in
some setting, made more precise later in this talk,
we are able to calculate all physical quantum states and to analyze
the quantum spectra of observables. We arrive at a family of quantum
gravity theories, where we have complete control over the phase space
of the underlying classical theory. This allows one to gain deep
insights into the relation between features of the quantum theory and
structures on the classical space-time.

The class of models considered comprises all 2D dilaton
gravity theories \cite{Odintsov} with a dilaton-dependent
coupling to a Yang-Mills (YM) field: \be L[g,\Phi,a] = \int_\CM d^2 x
\sqrt{|\det g|} \left[U(\Phi) R + V(\Phi) + W(\Phi) \6_\m \Phi \6^\m
\Phi + K(\Phi) tr( F_{\m \n} F^{\m \n} ) \right] \, .  \eel dil 
Here $g$ is the metric on the 2D space-time $\CM$, $\Phi$ is the dilaton
field, $F=da + a \wedge a$ is the curvature of the YM-connection $a$,
and $U,V,W,K$ are some basically arbitrary functions specifying the model.
In what will follow the gravity part can be extended 
further so as to include also terms giving rise to non-trivial torsion 
\cite{CQG}. 

Our first important observation is that, in a first order formalism,
the above gravity-Yang-Mills
systems may be reformulated as Poisson-$\s$-models
\cite{PSM}: \be L= \int_\CM A_i \wedge dX^i + \2 \CP^{ij}(X(x)) A_i
\wedge A_j \, .  \eel psact Here the $A_i$ and $X^i$ are a multiplet
of one-forms and functions on $\CM$, respectively, and $\CP^{ij}$ is
an antisymmetric $X^i$-dependent matrix, determined by $U,V,W,K$, satisfying
the crucial equality 
$\CP^{il}\,\6\CP^{jk}/\partial X^l  + cycl(i,j,k) = 0$.
Up to appropriate dilaton-dependent prefactors, $A_i$ collects
zweibein, spin-connection, and $a$, while, in the torsion free case,
the $X^i$ comprise two Lagrange multipliers enforcing torsion zero, the
dilaton field $\Phi$, and momenta associated to the YM-connection \cite{CQG}. 
The identity satisfied by
$\CP^{ij}$ turns out to equip the manifold $\CN$ spanned by the $X^i$ 
with a {\it Poisson bracket}
$\{ X^i,X^j \}_\CN := \CP^{ij}(X)$. 
This is the reason for calling the two-dimensional topological field
theories given by (\ref{psact}) 
Poisson-$\s$-models.  The analysis of
$(1+1)$-gravity-Yang-Mills systems is greatly facilitated by the
reformulation.\footnote{The situation closely resembles
  (2+1)-dimensional Einstein gravity, which can be reformulated as an
  $ISO(2,1)$-Chern-Simons theory; however, now there is a whole family of
  such gravity theories and the reformulation
  as a 
  topological field theory is of the form of an ordinary non-abelian 
gauge theory  only in exceptional cases.}
 
In particular, using the formalism provided by Poisson-$\s$-models
(cf.\ Part I of \cite{CQG} or \cite{PSM}), the classical field
equations can be brought into an extremely simple form locally. They give
rise to a (1 + rank(gauge group))-parameter family of gauge inequivalent
solutions, labelled by a mass parameter and Yang-Mills charges associated to
the Casimirs of the gauge group. The corresponding local expression
for the metric depends on the mass parameter $M$ and the charge $q$ of the
quadratic Casimir only and can be brought into the form 
\be
  g=2drdv+h_{M,q}(r)dv^2 \, ,
\eel locsol
where the family of functions $h_{M,q}$,
labelled by $M$ and $q$,
is determined by
the potentials $U,V,W,K$, characterizing the model (cf.\ Part I of
\cite{CQG}); e.g.\ for $U\equiv\Phi$, $W\equiv K\equiv 0$ one has 
$h=-\int V(r)dr +M$.

In general the local expression \rz locsol for the metric does not represent 
a complete space-time. To find the global solutions we have to construct
maximal extensions. Restricting ourselves in a first step to
(simply connected) universal covering solutions,
this may be done using the building block principle developed in
Part II of \cite{CQG}. Typically, this leads to black-hole configurations
such as depicted below.
\begin{figure}[htb]
\begin{center}
\leavevmode
\epsfxsize 9 cm \epsfbox{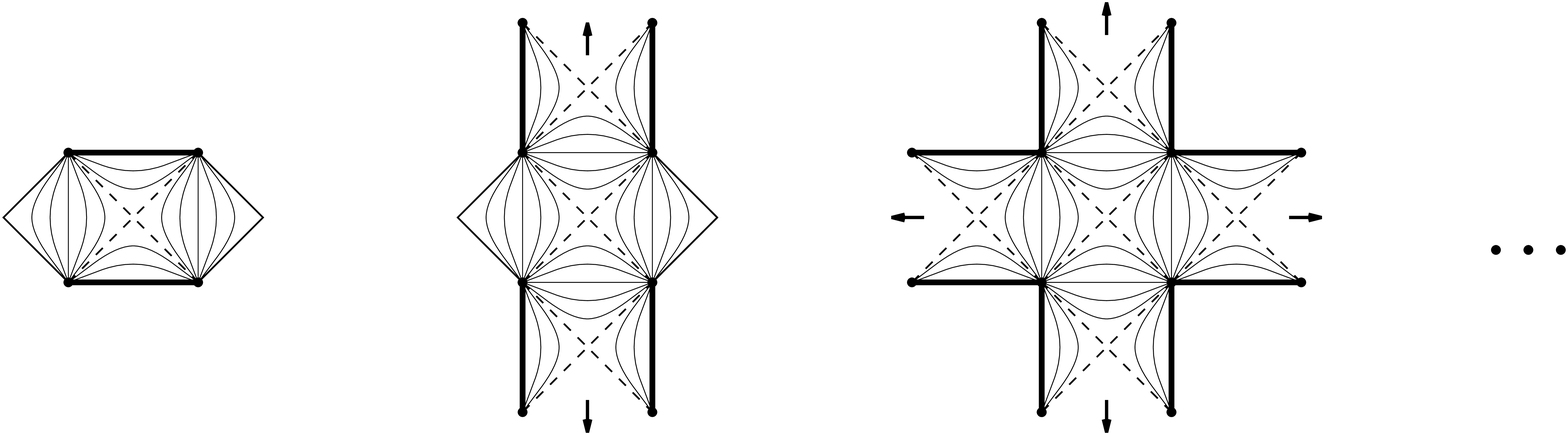}
\end{center}
\begin{quote}
{\bf Figure:} {\small Possible Penrose diagrams for an increasing number of
zeros of $h$ in \rz locsol .}
\end{quote}
\end{figure} 
Any solution with non-trivial first homotopy can then
be obtained in a second step by dividing out some discrete subgroup of the
isometry group from the universal covering. (Note that the isometries of the
metric extend to symmetries of the entire solution).
Restricting our attention to orientable and time-orientable
manifolds, the isometry group takes the simple form 
of a {\em direct\/} product
$\dR\times{\cal F}(k)$, where $\dR$ corresponds to transformations generated
by the Killing field $\6/\6 v$ and
${\cal F}(k)$ is a {\em free\/} group of rank $k$ with
\be
  k= \# \mbox{simple zeros of $h$} + 2 \# \mbox{multiple zeros of $h$} - 1 \,.
\ee
For a given universal covering
the diffeomorphism inequivalent maximally extended space-time solutions are in
one-to-one correspondence with the conjugacy classes of the
subgroups of this isometry group. For the above group they can be determined
explicitly and, in particular, for $k > 1$ smooth solutions on
two-surfaces of arbitrary genus with an arbitrary nonzero number of
punctures can be obtained! The full space of gauge inequivalent solutions
of (\ref{dil}) on such a two-surface 
has dimension $\left(\mbox{rank}(\pi_1(\CM))+1\right)$ times
$\left(\mbox{rank}(\mbox{gauge group})+1\right)$. For more details cf.\
Part III of \cite{CQG}.

To quantize the model \rz psact we choose a Hamiltonian Dirac approach. This
restricts $\CM$ to be of the form $\S \times \dR$ and we choose $\S =
S^1$.  As (\ref{psact}) is in first order form already, the Hamiltonian
structure may be read off easily: 
Denoting coordinates on $S^1 \times \dR$ by $(r,t)$, 
the $r$-components of the one-form valued
$A_i$-variables serve as the momenta for the fields $X^i(r)$, while
their $t$-components give rise to the set of first class constraints
$G^i(r) \equiv X^i(r)' + \CP^{ij} A_{j r} \approx 0$. On the quantum
level the latter are imposed as operator conditions selecting the
physical wave functions.

The solution to the quantum constraints of the field theory is
intimately related to the geometry of the target space, recognized
above as the one of a `Poisson manifold' $\CN$. A Poisson structure
$\CP$ on a manifold $\CN$ naturally induces a foliation of the
manifold into symplectic leaves, each of which carries a symplectic
structure $\O$.\footnote{The matrix $\CP^{ij}$ determined from
  (\ref{dil}) is degenerate; only by restriction to one of the
  above-mentioned submanifolds, the integral surfaces of the vector
  fields $\CP^{ij} \6/\6 X^j$, $\CP^{ij}$ becomes non-degenerate with
  inverse $\O$. The symplectic leaves may be characterized by
  constant values of functions on $\CN$ which
  may be identified with the mass parameter $M$ and the 
  YM-charges characterizing the local classical solutions.} 
These symplectic submanifolds may be regarded as
classical mechanical systems.  If their second homology is trivial, then
the mechanical systems can be quantized. If the symplectic leaves
have non-trivial second homology, then the quantization is possible only if
the symplectic form is integral, i.e.\ $\int_\s \O = 2\pi n \hbar, \,
\forall \s \in H_2(\mbox{leaf})$.  This condition will select a
discrete subset from the set of symplectic leaves with non-trivial
second homology.

One finds the following connection between the quantum theory
associated to the Poisson-$\s$-model and the quantum mechanical
systems defined on the symplectic leaves of $\CN$ \cite{PSM}: Physical
quantum wave functions of the Poisson-$\s$-model correspond to the
quantizable symplectic leaves of $\CN$. If the first homotopy of the
respective symplectic leaf is trivial, then there is precisely one
quantum state corresponding to it. Otherwise the quantum states
corresponding to a particular symplectic leaf are labelled by
`winding numbers' $l_1\ldots l_m$, where $m=\mbox{rank}\,\pi_1(\mbox{leaf})$.

Applying these results to the gravity-Yang-Mills systems, the 
quantum states can effectively be written as wave functions
depending on
\begin{itemize}
\item
\vspace{-.6em}
a continuous parameter $M$ (the mass parameter
characterizing the classical solutions)
\item
\vspace{-.6em}
a set of quantum numbers $j$ labelling the irreducible representations of the
gauge group
\item
\vspace{-.6em}
possibly further integer parameters $l_1\ldots l_m$, if the symplectic leaf
corresponding to the value of  $M$ has non-trivial
first homotopy. 
\end{itemize}

\vspace{-.6em}
The result presented above holds for Minkowskian signature of the space-time
topology. In the case of Euclidean space-time the symplectic leaves will have
a different topology. For some
models (e.g.\ deSitter gravity) the mass spectrum becomes
purely discrete, for others it remains continuous (e.g.\ spherically
reduced gravity), while in general the spectrum of the mass operator $M$
will be a mixture of both of these scenarios.

 
\end{document}